\documentclass[aps,preprint]{revtex4}%
\usepackage{amsfonts}
\usepackage{amsmath}
\usepackage{amssymb}
\usepackage{graphicx}%
\begin{document}
\title{Domain wall dynamics in an optical Kerr cavity}
\author{V\'{\i}ctor J. S\'{a}nchez--Morcillo, V\'{\i}ctor Espinosa, Isabel
P\'{e}rez--Arjona, }
\affiliation{Departament de Fisica Aplicada, Escuela Polit\'{e}cnica Superior de Gandia,
Universidad Polit\'{e}cnica de Valencia, Ctra. Natzaret-Oliva S/N, 46730-Grao
de Gandia, Spain}
\author{Fernando Silva, Germ\'{a}n J. de Valc\'{a}rcel and Eugenio Rold\'{a}n}
\affiliation{Departament d'\`{O}ptica, Universitat de Val\`{e}ncia, Dr. Moliner 50,
46100-Burjassot, Spain}

\begin{abstract}
An anisotropic (dichroic) optical cavity containing a self-focusing Kerr
medium is shown to display a bifurcation between static --Ising-- and moving
--Bloch-- domain walls, the so-called nonequilibrium Ising--Bloch transition
(NIB). Bloch walls can show regular or irregular temporal behaviour, in
particular, bursting and spiking. These phenomena are interpreted in terms of
the spatio-temporal dynamics of the extended patterns connected by the wall,
which display complex dynamical behaviour as well. Domain wall interaction,
including the formation of bound states is also addressed.

\end{abstract}
\maketitle

\section{Introduction}

Domain walls are localized structures typical of spatially extended systems
with broken phase invariance, where two or more homogeneous states with
different phases occupy different patches, the walls being defects connecting
two of such states or domains. Two different types of walls can exist in
systems described by a complex order parameter, namely Ising and Bloch walls,
which differ in the way the phase changes between the two domains as the wall
is crossed: An Ising wall is characterized by a discontinuous variation of the
field phase across the wall, whereas in a Bloch wall the phase angle rotates
continuously accross the wall. As this rotation can occur in two different
senses Bloch walls are chiral. (Contrarily, the Ising wall is not chiral.)
Alternatively, the order parameter is null at the core of an Ising wall, while
in a Bloch wall the order parameter does not vanish at any point. For this
reason, Ising and Bloch walls are often referred to as dark and grey solitons,
respectively. When the system dynamics does not derive from a potential other
crucial difference between Ising and Bloch walls refers to their dynamics:
Ising walls are stationary (static), while Bloch walls move with a velocity
related to their chirality.

Both types of domain walls may exist in different parameter regions, and in
this case they bifurcate one into another via a nonequilibrium Ising-Bloch
(IB) transition \cite{CoulletPRL,Michaelis}, which can be interpreted as a
bifurcation of the wall chirality \cite{CoulletPRL}. The IB transition has
been found in systems of very different nature, such as nematic liquid
crystals \cite{FrischPRL} or reaction--diffusion systems \cite{HagbergPRE}. In
the optical context, the phenomenon has been predicted to occur in type I
\cite{Perez04} and type II \cite{MaxiOL} optical parametric oscillators; from
the experimental point of view it is to be mentioned a related study
\cite{Larionova04} where Larionova \textit{et al }have analysed the dynamics
of two-dimensional phase domains in a photorefractive oscillator in a
degenerate four--wave mixing configuration:\textit{ }Within the same (closed)
wall they found both Bloch--type and Ising--type segments. They did not
observe, however, the nonequilibrium IB transition, probably because of the 2D
character of their system, which complicates the dynamics of domain walls with
curvature effects. In fact the first experimental observation of the
nonequilibrium IB transition in optics has been confirmed very recently in the
(transversely) one--dimensional version of the same device \cite{Esteban05}.
The possibility of domain walls as stable states has a particular interest in
nonlinear optical systems, given the potential use of localized structures in
all--optical signal processing. In nonlinear optical cavities, the structures
develop in the transverse plane, perpendicular to the resonator axis, and can
be controlled by external parameters \cite{PerezOpEx}.

Optical systems with a Kerr nonlinearity have been shown to exhibit a rich
spatio--temporal dynamics, see \cite{Arecchi99} and references therein for
details. Concerning the nonequilibrium Ising--Bloch transition, the Kerr
cavity, consisting of an optical resonator filled with a Kerr medium and
driven by a external coherent field, is a good candidate for exhibiting it, as
this system presents the basic requirement of broken phase invariance. The
original model proposed by Lugiato and Lefever \cite{LL87} was later extended
in \cite{Geddes94} to include the vector character of the light fields, what
allows to describe instabilities of the light polarization state. This work
was extended in \cite{Hoyuelos98} where the case of elliptically polarized
input was considered and in \cite{Gallego00} where two dimensional domain
walls in the form of dark-ring solitons were studied.

Recently these works were generalized by considering the possibility of
dichroism and/or birefringence in the optical cavity for the two linear
polarization components \cite{Sanchez00a,Sanchez00b,Perez01}. It was shown
that large enough cavity anisotropy or birefringence substantially modifies
the dynamics of the system. In particular, they allow for the polarization
instability to rule pattern formation in self-focusing Kerr cavities,
something that does not occur in isotropic cavities (without birefringence or
dichroism) \cite{Geddes94}. Moreover, as the polarization instability can be
subcritical for large enough dichroism or birefringence
\cite{Sanchez00b,Perez01} bright cavity solitons can exist, which in this case
are polarization solitons \cite{Sanchez00b,Perez01}. In \cite{Sanchez00b} it
was also shown that in the limit of large anisotropy (when the losses of the
two polarization components are very different in magnitude), the dynamics of
the system can be described by a single order parameter obeying a universal
equation, namely the parametrically--driven, damped nonlinear Schr\"{o}dinger
equation (PDNLSE). For our purposes the recent prediction \cite{deValcarcel02}
that the PDNLSE contains an IB transition and that Ising and Bloch domain
walls can connect either spatially uniform or patterned states is of special
relevance. This equation has been derived in different contexts (see
\cite{deValcarcel02}). In particular, in the optical context, the PDNLSE has
been shown to describe (apart from the anisotropic Kerr resonator
\cite{Sanchez00b}) the degenerate optical parametric oscillator
\cite{Longhi,Trillo97} and optical fibre loops with parametric amplification
\cite{Kath}.

However, despite the generality of the results obtained in
\cite{deValcarcel02}, their applicability to the Kerr cavity is strictly valid
only in the case of strong anisotropy, a very restrictive condition. In this
paper we consider the more realistic case of a Kerr cavity with moderate
anisotropy as discussed in Sec. \ref{model}, where the symmetries of the model
and the feasibility of domain walls are discussed. The pattern forming
properties of this system are investigated in Sec. \ref{instabilities}. Ising
and Bloch walls are then shown to be solutions of the system and the IB
transition between them is characterized in Sec. \ref{ib}. For small
detunings, the usual scenario of IB transition and Bloch wall dynamics is
found. For larger detunings, the dynamics of domain walls shows novel features
typical of excitable systems, such as spiking and bursting during the domain
wall evolution, as discussed in Sec. \ref{spiking}. The collisions between
walls and the formation of bound states is preliminary considered in Sec.
\ref{bound}. Finally the main conclusions of the work are highlighted in Sec.
\ref{conclusions}.

\section{Model and homogeneous solutions}

\label{model}The system considered in this paper is an optical resonator with
plane mirrors filled with an isotropic $\chi^{\left(  3\right)  }$ medium of
self-focusing type and driven by a spatially homogeneous linearly polarized
coherent field of amplitude $E$ (taken real without loss of generality) that
propagates along the resonator axis $z$. The resonator is anisotropic
(dichroic),\textit{\ i.e.}, the two intracavity field polarization components
$A_{0}$ and $A_{1}$ (parallel and orthogonal to the input field, respectively)
experience different linear losses (with associated cavity linewidths
$\gamma_{0}$ and $\gamma_{1}$ respectively). In order to avoid curvature
effects that strongly influence domain wall dynamics, we assume a transversely
1D problem (that can be experimentally achieved with a slab waveguide geometry
that confines the fields along one transverse dimension (say $y$), or with
rectangular slits placed in appropriate planes \cite{Estebanrolls}). The
adimensional model equations for such a system in the mean field limit read
\cite{Sanchez00b,Perez01}:%
\begin{align}
\partial_{t}A_{0}  &  =-\left(  \gamma+i\Delta_{0}\right)  A_{0}+i\left(
|A_{0}|^{2}A_{0}+\mathcal{A}|A_{1}|^{2}A_{0}+\tfrac{\mathcal{B}}{2}A_{1}%
^{2}A_{0}^{\ast}\right)  +i\partial_{x}^{2}A_{0}+\gamma E,\label{modelo1}\\
\partial_{t}A_{1}  &  =-\left(  1+i\Delta_{1}\right)  A_{1}+i\left(
|A_{1}|^{2}A_{1}+\mathcal{A}|A_{0}|^{2}A_{1}+\tfrac{\mathcal{B}}{2}A_{0}%
^{2}A_{1}^{\ast}\right)  +i\partial_{x}^{2}A_{1}, \label{modelo2}%
\end{align}
where $\Delta_{0}=\left(  \omega_{0}-\omega\right)  /\gamma_{1}$ and
$\Delta_{1}=\left(  \omega_{1}-\omega\right)  /\gamma_{1}$ are normalized
cavity detunings ($\omega$ is the angular frequency of the input field and
$\omega_{0,1}$ are the frequencies of the cavity longitudinal modes with
polarization parallel and orthogonal to the input closest to $\omega$),
$\gamma=\gamma_{0}/\gamma_{1}$ is the cavity anisotropy parameter;
$\partial_{x}^{2}$ accounts for diffraction (the transverse spatial coordinate
$x$ is normalized to the diffraction coefficient) and $t$ is time normalized
to $\gamma_{1}^{-1}$. Finally, $\mathcal{A}$ and $\mathcal{B}$ are the Maker
and Terhune coefficients, which verify $\mathcal{A}+\frac{\mathcal{B}}{2}=1$
for isotropic media \cite{Boyd}, which we consider. For details on the
normalizations see \cite{Sanchez00a}.

In Refs. \cite{Geddes94,Hoyuelos98,Gallego00}, the pattern formation
properties of this model have been studied for $\gamma=1$ and $\Delta
_{1}=\Delta_{0}$. We extended the model to the anisotropic cavity ($\gamma
\neq1$) first considering the plane--wave model (no diffraction)
\cite{Sanchez00a} and then considering pattern formation \cite{Sanchez00b},
where we concentrated in the limit of large cavity anisotropy and demonstrated
that the PDNLSE describes the system in this limit. Later, in \cite{Perez01}
we studied patern formation and localized structures due to birefringence
($\Delta_{0}\neq\Delta_{1}$).

In the following, the case $\Delta_{0}=\Delta_{1}\equiv\Delta$ (no
birefringence) is considered. Also we restrict the study to the case when the
nonlinear material is a liquid (e.g., \textrm{CS}$_{2}$), for which
$\mathcal{A}=1/4$ and $\mathcal{B}=3/2$ \cite{Boyd}. This makes that our
results could apply, e.g., to cells of liquid crystal in the isotropic phase
(nematic liquid crystals, which have a much larger nonlinearity, are not
covered by our analysis as they are \textit{anisotropic} nonlinear media).

Most relevant for our study are the symmetries supported by Eqs.
(\ref{modelo1},\ref{modelo2}). In particular the term $\gamma E$ in Eq.
(\ref{modelo1}) completely breaks the phase symmetry of field $A_{0}$.
Nevertheless, the model still supports the discrete symmetry $\left(
A_{0},A_{1}\right)  \rightarrow\left(  A_{0},-A_{1}\right)  $. (Note that the
four-wave mixing term -the one multiplied by $\tfrac{\mathcal{B}}{2}$ in Eq.
(\ref{modelo2})- breaks the continuous phase symmetry.) This symmetry means
that whenever a state $\left(  A_{0}\left(  x,t\right)  ,A_{1}\left(
x,t\right)  \right)  $ is a solution of the system, another solution $\left(
A_{0}\left(  x,t\right)  ,-A_{1}\left(  x,t\right)  \right)  $ exists as well
that is dynamically equivalent (has the same dynamical properties like
stability, etc.) to the former. As $A_{0}$ and $A_{1}$ correspond to the two
orthogonal components of the light electric field vector, the above symmetry
relates two equivalent solutions having opposite helicity, appart from a
different polarization ellipse orientation. This symmetry thus opens the
possibility of exciting domain walls that join asymptotically two of such
symmetric states. On the other hand, the reflection $\left(  x\rightarrow
-x\right)  $ and translation $\left(  x\rightarrow x+x_{0}\right)  $
invariances of the problem imply that if a Bloch wall of given chirality
exists, another, equivalent one of opposite chirality also exists, and both
move in opposite directions \cite{CoulletPRL,deValcarcel02,Esteban05}.

The existence conditions and dynamic behaviour of domain walls, the main
subject of this article, is strongly related with the stability properties of
the homogeneous solutions connected by the walls. In the system described by
Eqs. (\ref{modelo1}) and (\ref{modelo2}), these solutions have been analyzed
in \cite{Sanchez00a,Perez01}, and we review here the main results.

According to the polarization state of the intracavity field two kinds of
steady homogeneous solutions are possible: (i) The \textit{linearly polarized
state}, with intensities $I_{1}\equiv\left\vert A_{1}^{2}\right\vert =0$ and
$I_{0}\equiv\left\vert A_{0}^{2}\right\vert $ given by the solutions of
\begin{equation}
\gamma^{2}E^{2}=\left[  \gamma^{2}+\left(  \Delta-I_{0}\right)  ^{2}\right]
I_{0}, \label{linhom}%
\end{equation}
and (ii) the \textit{elliptically polarized state}, with intensities
determined by
\begin{align}
\gamma^{2}E^{2}I_{0}  &  =\left(  I_{1}+\gamma I_{0}\right)  ^{2}+\left[
\left(  \Delta-I_{1}\right)  I_{1}-\left(  \Delta-I_{0}\right)  I_{0}\right]
^{2},\label{pol.hom.1}\\
I_{1}  &  =\Delta-\mathcal{A}I_{0}\pm\sqrt{\left(  \frac{\mathcal{B}}{2}%
I_{0}\right)  ^{2}-1}. \label{pol.hom.2}%
\end{align}
The linearly polarized solution Eq. (\ref{linhom}) shows a multivalued
character when $\Delta>\sqrt{3}\gamma$ \cite{Sanchez00a}. The elliptically
polarized solution Eq. (\ref{pol.hom.1}) was analyzed in detail in
\cite{Sanchez00a} and can be a multivalued function as well.

The cavity anisotropy parameter $\gamma$ plays an important role on the
character of the polarization instability (i.e., the bifurcation affecting the
linearly polarized solution that sets the onset of eliptically polarized
emission), as shown in \cite{Sanchez00a,Perez01}. In particular the
polarization instability can become subcritical for $\gamma>2$ (the exact
value at which this occurs depends on the detuning value) whenever
$\Delta>1/3$. In this case bright cavity solitons can be supported by the
system at low pumpings \cite{Sanchez00b}. More important for the present study
is the fact that for large $\gamma$ the anisotropic Kerr cavity model can be
reduced to a PDNLSE, which exhibits an IB transition \cite{deValcarcel02}.
This implies that the anisotropic Kerr cavity will exhibit the same phenomenon
for large enough $\gamma$. In order to test the universality of this
phenomenon we investigate here a cavity with moderate anisotropy and take
$\gamma=3.5$ for definiteness. We note that for this value of $\gamma$ the
model cannot be rigorously reduced to a PDNLSE and then some extra features
can be expected.

\section{The pattern forming instabilities}

\label{instabilities}The stability of the linearly polarized solution
(\ref{linhom}) against space-dependent perturbations was analyzed in
\cite{Sanchez00b,Perez01}, where analytical expressions for the different
boundaries and wavenumber of the emerging patterns were obtained. In Fig. 1
the stability of the different solutions is shown on the plane $E-\Delta$,
which are the only free parameters. In the figure, the linearly polarized
solution (\ref{linhom}) is stable below curve $\mathrm{E}_{1}$, which
corresponds to the polarization instability. Above this curve the linearly
polarized state is no more a stable solution and gives rise to the
elliptically polarized solution (\ref{pol.hom.1},\ref{pol.hom.2}). In its
turn, this last solution exists above line $\mathrm{E}_{2}$, i.e., there is a
domain of coexistence between the linearly and elliptically polarized
solutions, marked in the figure as BS, between lines $\mathrm{E}_{1}$ and
$\mathrm{E}_{2}$.

We consider now the existence of pattern forming instabilities of the
elliptically polarized solution (\ref{pol.hom.1},\ref{pol.hom.2}), which are
of relevance for the analysis of the dynamics of domain walls performed below.
Following the usual procedure, we consider perturbations of the homogeneous
solutions in the form $\delta A_{i}(x,t)=\delta A_{i}\,\exp\left(  \lambda
t+\mathrm{i}kx\right)  $, and Re$(\lambda)=0$ signals a bifurcation. Unlike
the case of the linearly polarized solution \cite{Sanchez00b,Perez01}, the
stability analysis is now quite involved and analytic expressions can not be
obtained. Instead, we perform a numerical analysis of the eigenvalues
$\lambda$ to determine the instability boundaries.

The pattern formation instability boundary affecting the elliptically
polarized solution (\ref{pol.hom.1},\ref{pol.hom.2}) corresponds to the
piecewise curve (with dashed and dotted segments) joining the points $a-e$.
Beyond this line (shadowed area), the elliptically polarized solution is
modulationaly unstable.

The complex form of this boundary follows from the dependence of the real part
of the largest eigenvalue with the wavenumber of the perturbation. As shown in
Fig. 2, the real part of the eigenvalue (solid line) evaluated for parameters
corresponding to point (b) in Fig. 1 shows two maxima at different
wavenumbers. Depending on the parameter setting the threshold is minimum for
the smallest wavenumber (dashed lines $b-c$ and $d-e$) or for the largest
wavenumber (dotted lines $a-b$ and $c-d$). Consequently, the points denoted by
$b$, $c$ and $d$ in the figure correspond to codimension-2 points, where the
instability is reached at two different wavenumbers simultaneously.
Furthermore, the imaginary part of the eigenvalue is null in the first cases
(dashed lines $b-c$ and $d-e$), thus corresponding to the emergence of
stationary patterns, but non null in the second cases (dotted lines $a-b$ and
$c-d$), see Fig. 2, thus corresponding to a Hopf bifurcation that gives rise
to the appearance of dynamic patterns whose amplitude oscillates in time.
Finally, as stated, the region marked with \textrm{BS} in Fig. 1 denotes the
domain of coexistence between linearly and elliptically polarized states. When
this coexistence is bistable and one of the states (the elliptically polarized
state in this case) is spatially modulated (shadowed region), bright cavity
solitons can be also excited \cite{Sanchez00b,Perez01}.

\section{The Ising-Bloch transition}

\label{ib}We performed the numerical integration of Eqs. (\ref{modelo1}) and
(\ref{modelo2}) with a split--step algorithm, for $\mathcal{A}=1/4$,
$\mathcal{B}=3/2$, $\Delta_{0}=\Delta_{1}\equiv\Delta$, and $\gamma=3.5$.
Periodic boundary conditions were used that impose an even number of walls in
the transverse domain. A spatial grid of 2048 points was used and the temporal
step was lowered down to $\Delta t=0.001$ in order to otain $\Delta
t-$independent results. As an initial condition, two walls were placed
symmetrically with respect to the center of the integration window. Several
integration region lengths $L$ were investigated. Results shown here
correspond to the choice $L=20\sqrt{5}$, which ensured enough spatial
separation between the two walls in order to avoid their mutual interaction
during the initial stage of the evolution. For some parameter settings the
walls reached a static configuration after a transient, having fixed their
position across the transverse plane. In such cases it was assessed that walls
were Ising ones by verifying that the complex field $A_{1}$ was zero at the
wall core. For other parameter values walls reached a moving configuration and
were identified as Bloch walls as there was no point in the transverse plane
where the complex field $A_{1}$ was zero. Note that the following analysis of
the results refers to the cross polarized component $A_{1}$ that is the one in
which domain walls can be clearly identified as dark solitons
\cite{Sanchez00b} as walls join two domains where the values of $A_{1}$ have
opposite signs (phases) as discussed.

Figure 1 summarizes our numerical findings. For low values of both pump $E$
and detuning $\Delta$ (inside the region labeled "Ising", delimited by the
solid lines joining points $a-f-g-h-d-e$, and line $\mathrm{E}_{2}$) stable
Ising walls are found. These Ising walls connect homogeneous states in most of
this region, except in the thin white areas at the right of the dashed line
$d-e$ and the dotted line $c-d$, where they connect patterned states, in
agreement with the linear stability analysis discussed above. By increasing
the pump or the detuning from this region we observe IB transitions, marked
with solid lines $f-g-h-d-e$, where Ising walls are replaced by Bloch walls.

Beforme commenting the differences between the various Ising--Bloch transition
lines ($\mathrm{IB}_{1}$, $\mathrm{IB}_{2}$ and $\mathrm{IB}_{3}$), let us
comment about what happens in the small region above line $a-b$, where the
behaviour is somewhat anomalous. In the small domain between this line and the
dark--grey shadowed area marked as "\textrm{No walls}" the walls are unstable,
and the single pattern supported by the system are rolls. Then, strictly
speaking, the domain "\textrm{No walls}" extends until the line $a-b$. If we
have let this domain without including it in the "\textrm{No walls}" domain is
just because in it a very long transient behaviour in the dynamics of the
walls is observed until walls eventually dissapear and the system develops a
spatially periodic state. This behaviour (longest transient) is in contrast
with what happens in the "\textrm{No walls}" domain, where domain walls
dissapear sharply.

At the right of the point $f$, when increasing the pump, two different regimes
can be clearly distinguished, depending on detuning. For small detunings the
transition is denoted by $\mathrm{IB}_{1}$ in Fig. 1 (continuous line), and in
this case the Bloch walls (that exist inside the "triangle" $f-b-g$) connect
homogeneous states. For the chosen values of the parameters, this regime
exists up to $\Delta=1.75.$ For larger detunings, the transition is mainly
ruled by the pattern forming instability experienced by the domains joined by
the walls ($\mathrm{IB}_{2}$ in Fig. 1, represented by continuous curve). Note
that, for moderate detunings (in the center of the plot), the $\mathrm{IB}$
and pattern forming boundaries are nearly coincident. In this case the Bloch
walls always connect patterned states. Finally, for small pump but large
detunings, another region $\mathrm{IB}_{3}$ of Bloch walls is found (right
side of the curve $h-d-e$). The Bloch walls in this region move with a
extremelly small, random velocity, similarly to what happens in the PDNLSE
without diffusion or saturation terms \cite{deValcarcel02}.

An example of a Bloch wall corresponding to the transition $\mathrm{IB}_{1}$,
obtained for $\Delta=0.8$ and $E=3.25$, is shown in Fig. 3. In Fig. 3(a) the
intensity distribution in transverse space near the core of the wall is given,
and in Fig. 3(b) we show the corresponding phase--portrait (i.e., a plot of
the real versus the imaginary part of the field). Both pictures are
alternative representations where the Bloch character of the wall is
evidenced: The intensity at the wall core is small, but non null, Fig. 3(a),
and a smooth variation of the phase between the domains separated by the wall
is appreciated in the parametric plot in Fig. 3(b). For these parameters, the
homogeneous solutions which constitute the domains connected by the wall are
modulationally stable.

As stated in the introduction, an essential feature of Bloch walls in
nonvariational systems is that, contrary to Ising walls, they move in the
transverse plane, with a velocity that depends on the parameters. For a fixed
value of the detuning $\Delta=1$, the dependence of the wall velocity with the
pump is shown in Fig. 4. For this particular value of the detuning, in the
region where the background solutions are modulationally unstable, the walls
no more exist as commented, since spatial modulations of the background grow
and fill completely the transverse space leading eventually to a roll pattern.

Let us now describe the dynamic behaviour of Bloch walls.

\section{Bound states}

\label{bound}As stated in the previous section the use of periodic boundary
conditions forces that the number of domain walls that can exist within the
integration window be an even number, two is the minimum. On the other hand
Bloch walls move with a velocity whose sign depends on their chirality. Then
two situations are possible, namely, that the two Bloch walls have either the
same or different chiralities (notice that the \textit{sign} of the chirality
of the Bloch walls is fixed by the initial conditions). When the chiralities
are the same, the two Bloch walls move along parallel paths and no interaction
between them appears (at least when their movement is regular: When it is
highly irregular the chirality of the walls can change independently of each
other \cite{deValcarcel02}). Contrarily, when the chiralities are opposite the
paths followed by the two Bloch walls intersect and a collision occurs. As a
result a new localized structure may appear and a bound state is formed. This
bound state is a cavity soliton different from a wall. Although a detailed
study of these objects falls outide the scope of the present work, and without
trying to be exhaustive in their characterization, we just note that two
different behaviours of the new cavity soliton have been identified.

For small detunings (i.e. inside the triangle $f-b-g$ above the region
$\mathrm{IB}_{1}$ in Fig. 1), after the collision the positions of the walls
delimiting the cavity soliton perform small amplitude antiphase periodic
oscillations, as shown in Fig. 4; hence the cavity soliton performs a
breathing dynamics, remaining constant its position in the transverse plane.

For higher detunings (above the region $\mathrm{IB}_{2}$) bound states are
also formed as a result of the interaction, but their dynamics is different.
In these cases, after the collision one wall is dragged by the other and the
resulting bound state drifts with the velocity of one of the original Bloch
walls as shown in Fig. 5.

Finally, we note that near but below the boundary $a-b$, the walls do not form
bound states after the interaction, but instead bounce and exchange their
chiralities. Such behaviour is shown in Fig. 6.

Although it is difficult to determine the origin of these behaviours after a
collision is produced, the reason for these different behaviours is very
likely linked to the different dynamics of the patterned state that the walls
connect, whose complicated spatio--temporal dynamics depends on the parameter
set. In the next section, where we concentrate in the dynamic behaviour of
isolated Bloch walls, we give some clues on how the pattern dynamics affects
the wall behaviour.

\section{Bursting and spiking dynamics of isolated domain walls}

\label{spiking}In the small detuning regime, Bloch walls behave in a regular
manner and the motion occurs at a constant velocity as in the case shown in
Fig. 7. For higher detunings however, the wall dynamics shows features which
are characteristic of excitable systems. The wall motion is in fact regular
only close to the $\mathrm{IB}_{2}$ transition, which nearly coincides with
the pattern formation boundary, see Fig. 1. For higher pump values, wall
dynamics is characterized by an irregular behaviour of the wall position. We
report next several numerical examples of such irregular motion, obtained for
$E=3.5$ and different values of the detuning $\Delta$.

In Fig. 8, a bursting phenomenon (the appearance of almost periodic
oscillations during time intervals of arbitrary duration), is observed both in
the wall position, Fig. 8(a), and chirality, Fig. 8(b), for $\Delta=2.25$. (We
used the definition of chirality $\chi=\left.  \operatorname{Im}\left(
A_{1}^{\ast}\partial_{x}A_{1}\right)  \right\vert _{x=x_{0}}$
\cite{deValcarcel02} where $x_{0}$ denotes the point where the wall intensity
$\left\vert A_{1}\right\vert ^{2}$ is at its minimum). After experiencing
several bursts, the wall velocity turns to be constant again, but in this case
with a chirality of opposite sign with respect to the initial one. During the
steady motion, the chirality takes a very small value, $\chi\approx10^{-3}$,
and consequently the change in sign is not appreciated in Fig. 8(a), owing to
the scale imposed by the bursting events. However, this change is manifested
in Fig. 8(b) as a change in the sign of the wall velocity. The number of
bursts and final chirality depends in general on the parameter values and
initial conditions, such as the amplitude and relative position of the walls.

A slight increase in the detuning leads to a qualitatively different
behaviour, namely a spiking in the wall position. An example is shown in Fig.
9, obtained for $\Delta=2.5$. In this case the bursts appear periodically in
time, in the form of spikes. The intensity distribution of the Bloch wall
during the spiking regime is shown in Fig. 10. The wall connects two patterned
states with different spatial distributions. The pattern on the right side of
the wall in Fig. 10 is spatially harmonic, contains a single spatial
frequency, while the pattern of the left side of the wall is bi-periodic, and
both a fundamental and a small amplitude second spatial harmonic are present.
This particular structure of the domain walls has been observed in all the
numerical simulations in the irregular regime, and seems to be at the root of
the complex behaviour exhibited by the wall dynamics.

These numerical results suggest that the complex dynamic behaviour of Bloch
walls is related with secondary instabilities of the extended roll patterns
which form the domains at both sides of the wall. In particular, numerics show
that bursting and spiking of the wall position always develop in coincidence
with the appearance of a second spatial frequency in the intensity of the roll
pattern forming one of the domains. To check this statement, Eqs.
(\ref{modelo1}) and (\ref{modelo2}) were integrated in the absence of walls,
and the spatial distribution of roll patterns was studied. A summary of
results is given in Fig. 11, corresponding to roll patterns obtained for a
fixed value of the pump $E=3.5$ and for different detunings, in accordance
with Figs. 8--9. The distributions shown in Figs. 11(a)--(c) correspond, from
top to bottom inside each figure, to the intensity, the real and the imaginary
parts of the extended patterns. Figure 11(a) has been obtained for
$\Delta=1.5$, and the resulting intensity pattern is a stationary roll, with a
single spatial frequency. In Fig. 11(b), where $\Delta=2$, the intensity
pattern develops a second spatial frequency of small amplitude. This value of
detuning nearly corresponds to the onset of bursting phenomena of the wall.
Finally, in Fig. 11(c), for $\Delta=2.25,$ the resulting intensity pattern
shows the coexistence, in different spatial domains, of rolls with different
periodicity and equal phases (compare with Fig. 10 where a domain wall
separates oppositely phased patterns). Clearly, in this parameter region there
exists bistability between different spatial structures. This bistability
seems to be a requirement for the existence of irregular dynamics. It is
interesting to interpret these results in terms of the real and imaginary
parts of the patterns shown in the bottom parts of Figs. 11(a)--(c). The
second spatial frequency of the intensity pattern, and consequently the
irregular dynamical behaviour, appears when real and imaginary parts of the
pattern distributions cross, Fig. 11(b) and left part of Fig. 11(c).

\section{Conclusions}

\label{conclusions}We have studied the dynamics of domain walls in an
anisotropic optical Kerr cavity. Both Ising and Bloch walls, and the
transition between them, have been reported in the case of moderate cavity
anisotropy. The stability of the homogeneous solutions against pattern forming
instabilities has been also analyzed. These results show a complex scenario of
spatio-temporal evolution of patterns in this system. Domain wall dynamics is
shown to be related with the stability of the domain (background) solutions. A
numerical study shows the existence of different domains of behaviour,
depending on parameters. Besides the typical evolution of Bloch walls, with a
drift at nearly constant velocity, we have observed regimes in which the
behaviour of the Bloch wall parameters (position, velocity, chirality and
intensity at the core) is irregular, and analogous to that found in the
temporal dynamics of excitable systems. These regimes, namely bursting and
spiking, are reported for the first time in the case of domain walls in
optical cavities. Finally new cavity solitons formed by the interaction of two
domain walls have been identified and two dynamical regimes of wall collision
have been envisaged.

\section{Acknowledgments}

We gratefully acknowledge helpful comments by Pier Luigi Ramazza. This work
was financially supported by the Spanish Ministerio de Ciencia y
Tecnolog\'{\i}a and European Union FEDER, under\ projects BFM2002-04369-C04-01
and BFM2002-04369-C04-04.

\pagebreak

{\Large Figure Captions}

\vspace{1cm}

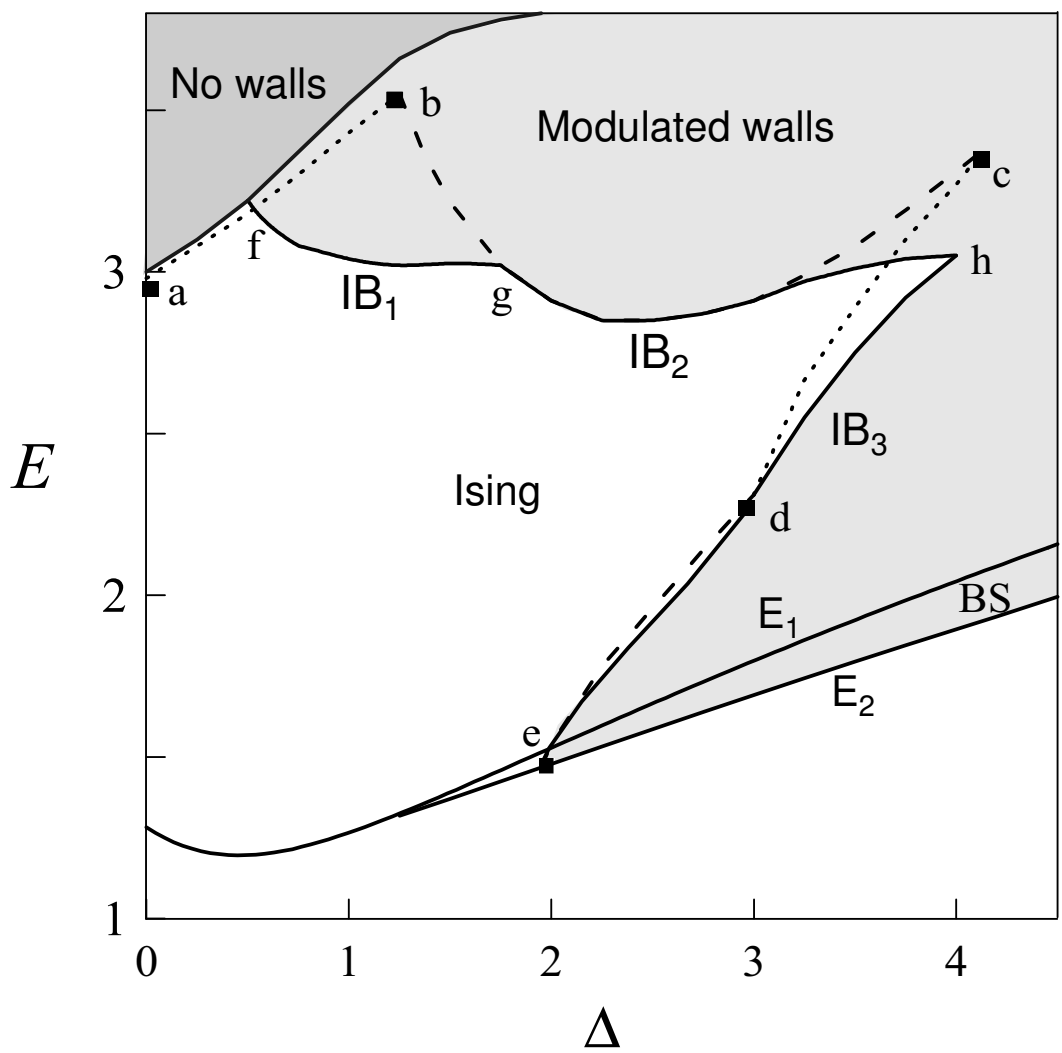
Figure 1. Bifurcation diagram of homogeneous solutions for $\mathcal{A}=1/4$,
$\mathcal{B}=3/2$, $\Delta_{0}=\Delta_{1}=\Delta$, and $\gamma=3.5$, together
with the boundaries of Ising-Bloch transitions. See text for details.

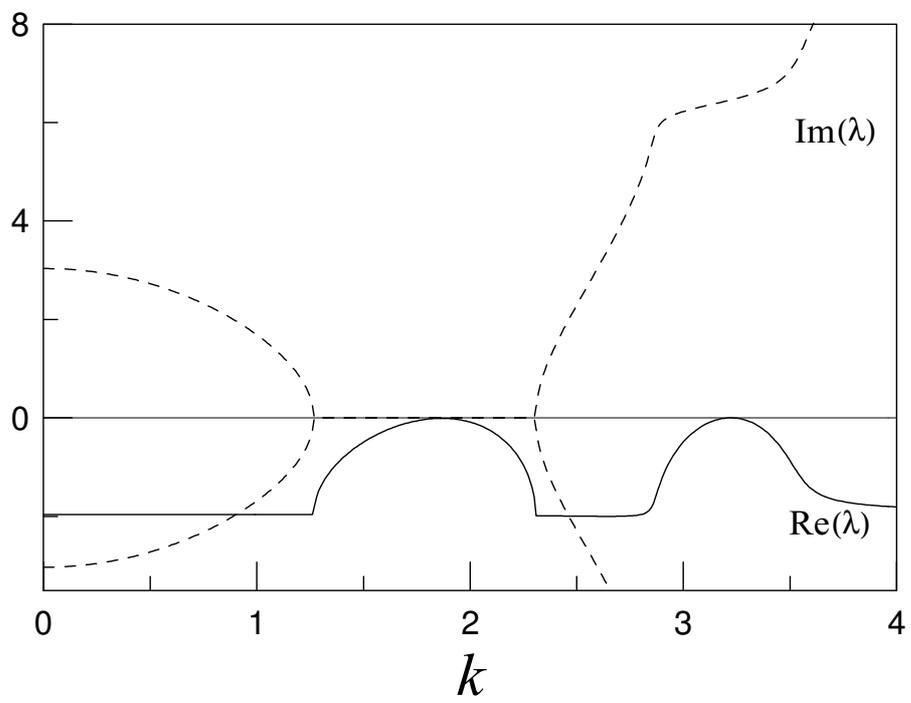
Figure 2. Growth rate of the perturbations of the homogeneous solutions as a
function of the spatial wavenumber, near a codimension-two point for
$\Delta=1.25$ and $E=3.5$. The rest of parameters as in Fig.1.

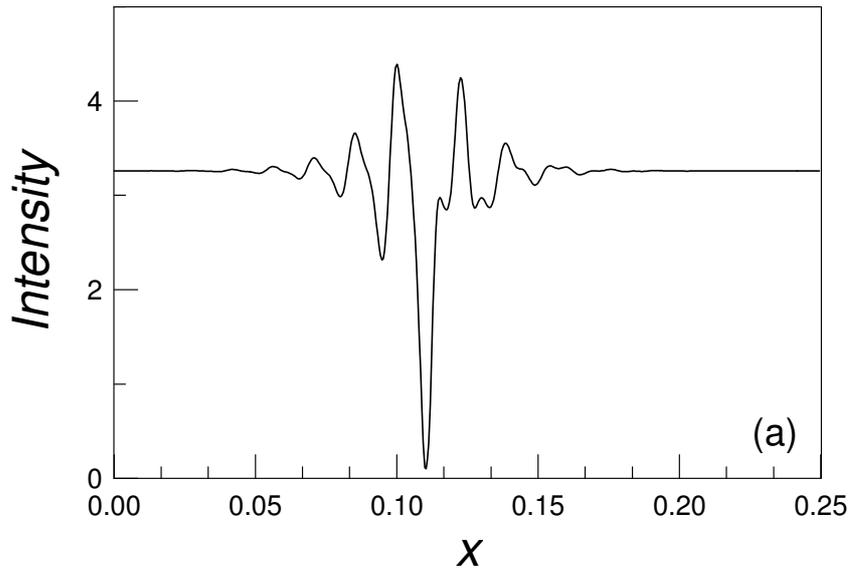
Figure 3. Example of a pair of Bloch walls numerically obtained for the
parameters $\Delta=0.8$ and $E=3.25$ (the rest of parameters as in Fig.1)
Intensity distribution (a) and parametric representation (b).

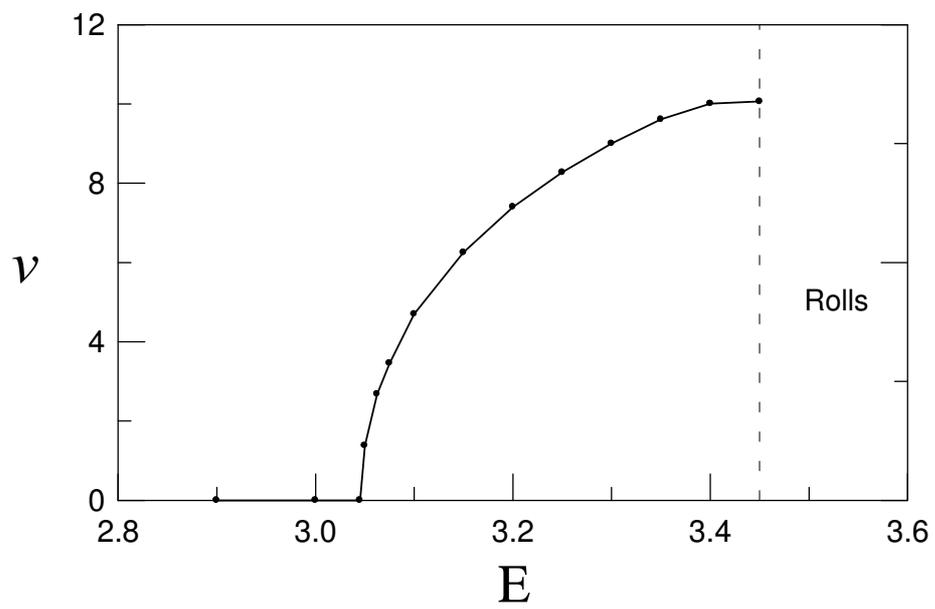
Figure 4. Bifurcation diagram of the velocity of the walls for $\Delta=1$ (the
rest of parameters as in Fig.1). Dashed line shows the boundary of the pattern
forming instability

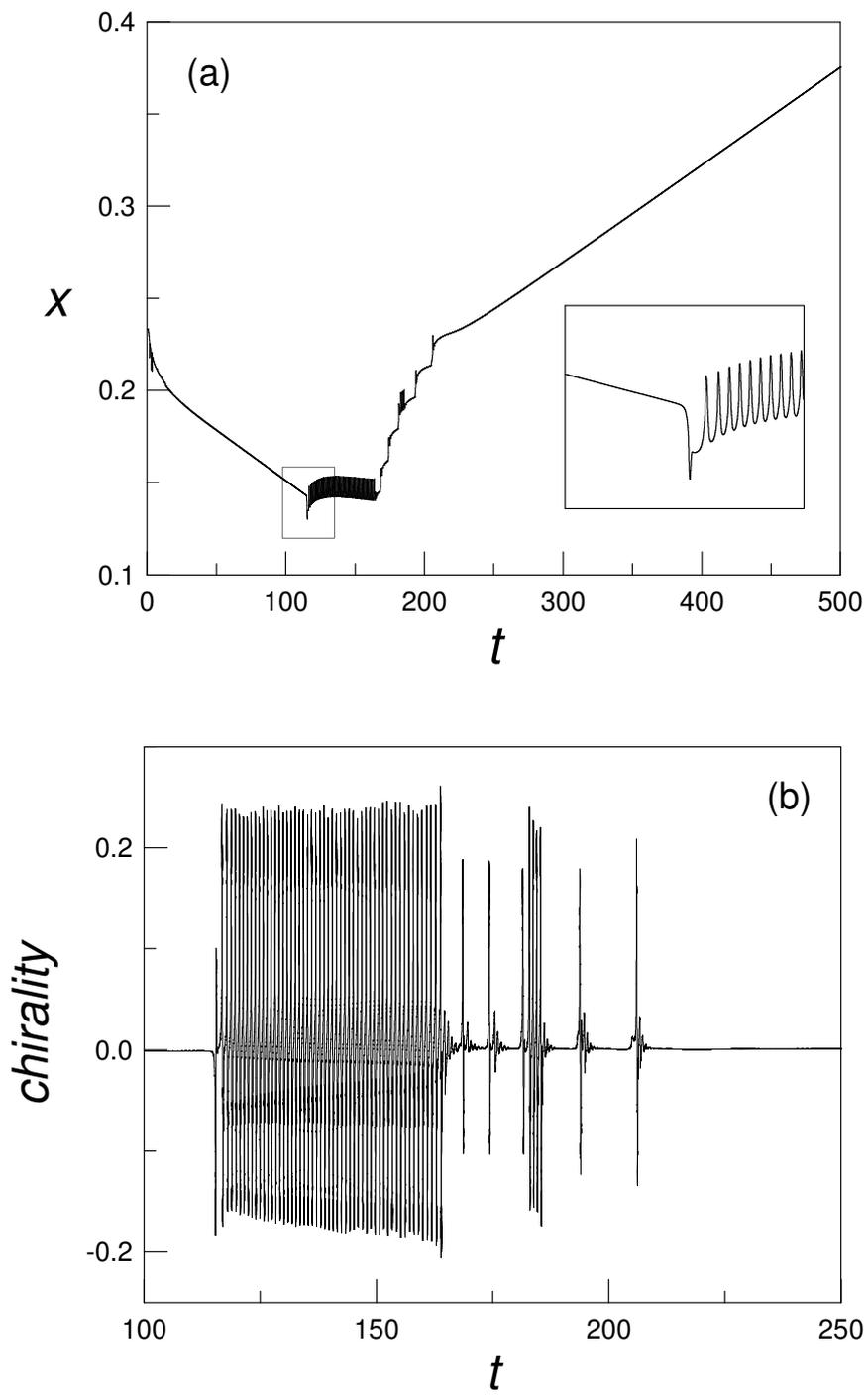
Figure 5. Bursting of the wall position (a) and chirality (b), obtained for
$\Delta=2.25$, $E=3.5$ (the rest of parameters as in Fig.1). Inset shows the
oscillations during the burst.

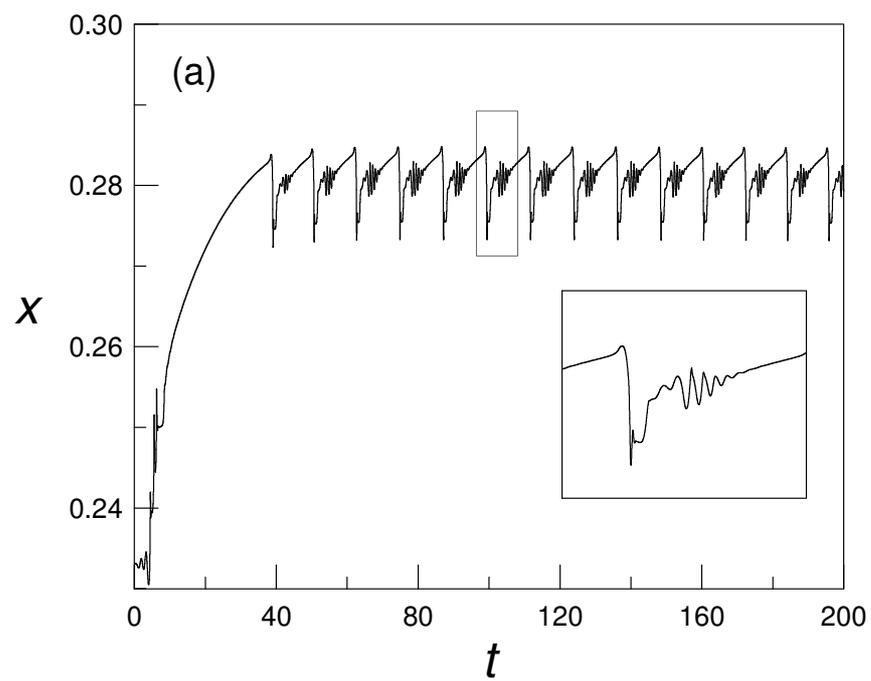
Figure 6. Spiking behaviour of the wall position (a) and chirality (b),
obtained for $\Delta=2.5$, $E=3.5$ (the rest of parameters as in Fig.1).

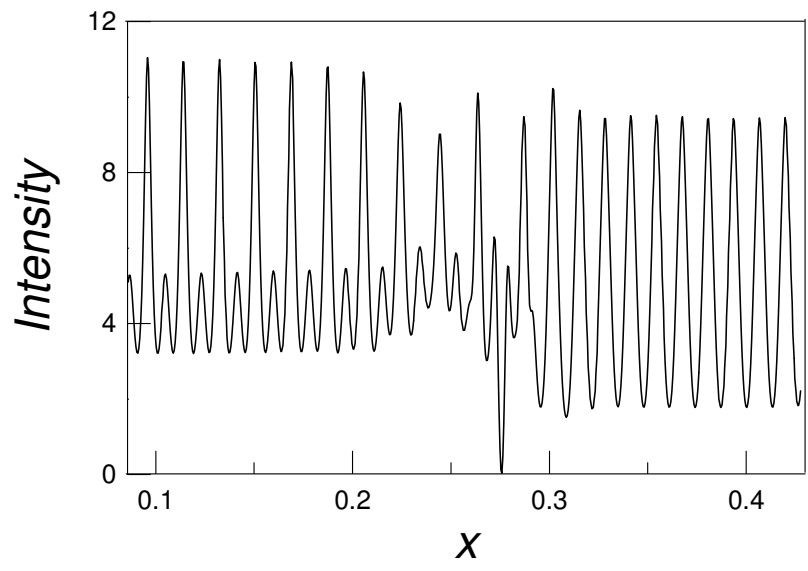
Figure 7. Intensity distribution near a Bloch wall corresponding to Fig.7.
Note that the wall connects periodic patterns with different spatial structure.

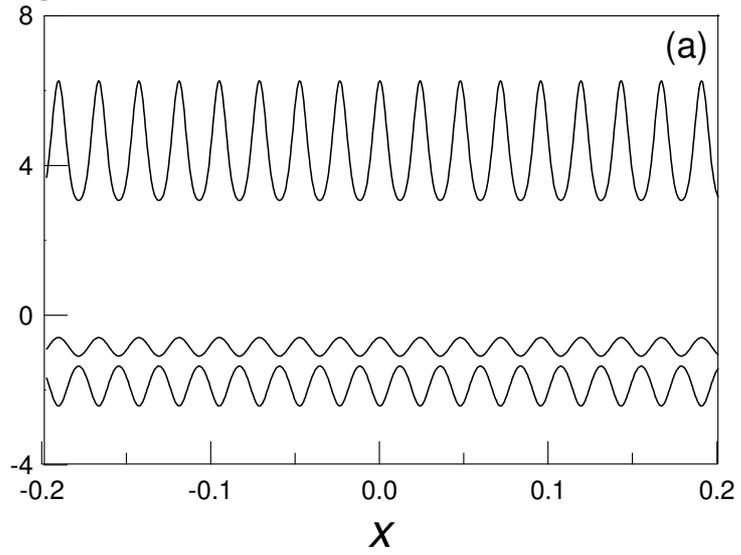
Figure 8. Roll patterns obtained at different detunings, for $E=3.5$ (the rest
of parameters as in Fig.1): (a) Stationary roll pattern with a single spatial
frequency ($\Delta=1.5$); (b) A second spatial frequency appears ($\Delta=2$);
(c) Bistability of roll patterns with different structure ($\Delta=2.25$).
Patterns in (b) and (c) are dynamic.

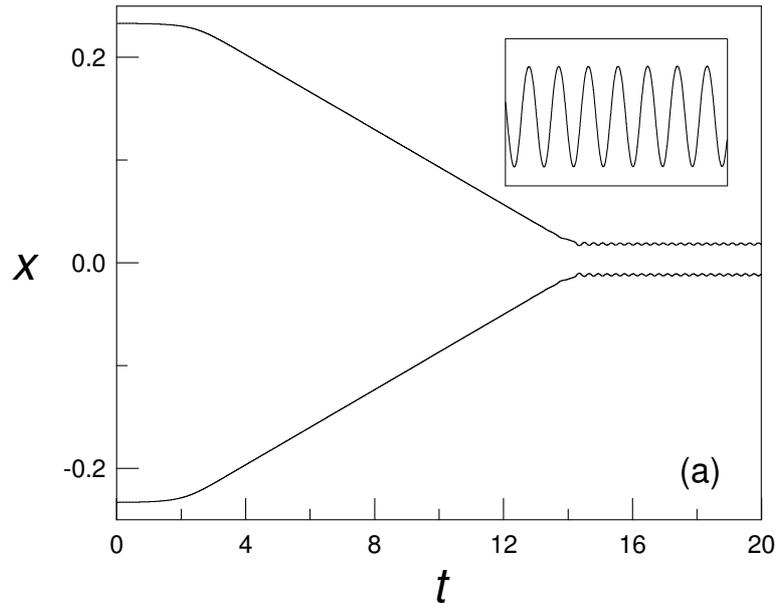
Figure 9. Interaction of two Bloch walls with opposite chirality for
$\Delta=1$ and $E=3.1$ (the rest of parameters as in Fig.1). Intensity (b) and
real part (c) distribution of the bound state. The inset shows the oscillatory
evolution of the wall position after the interaction

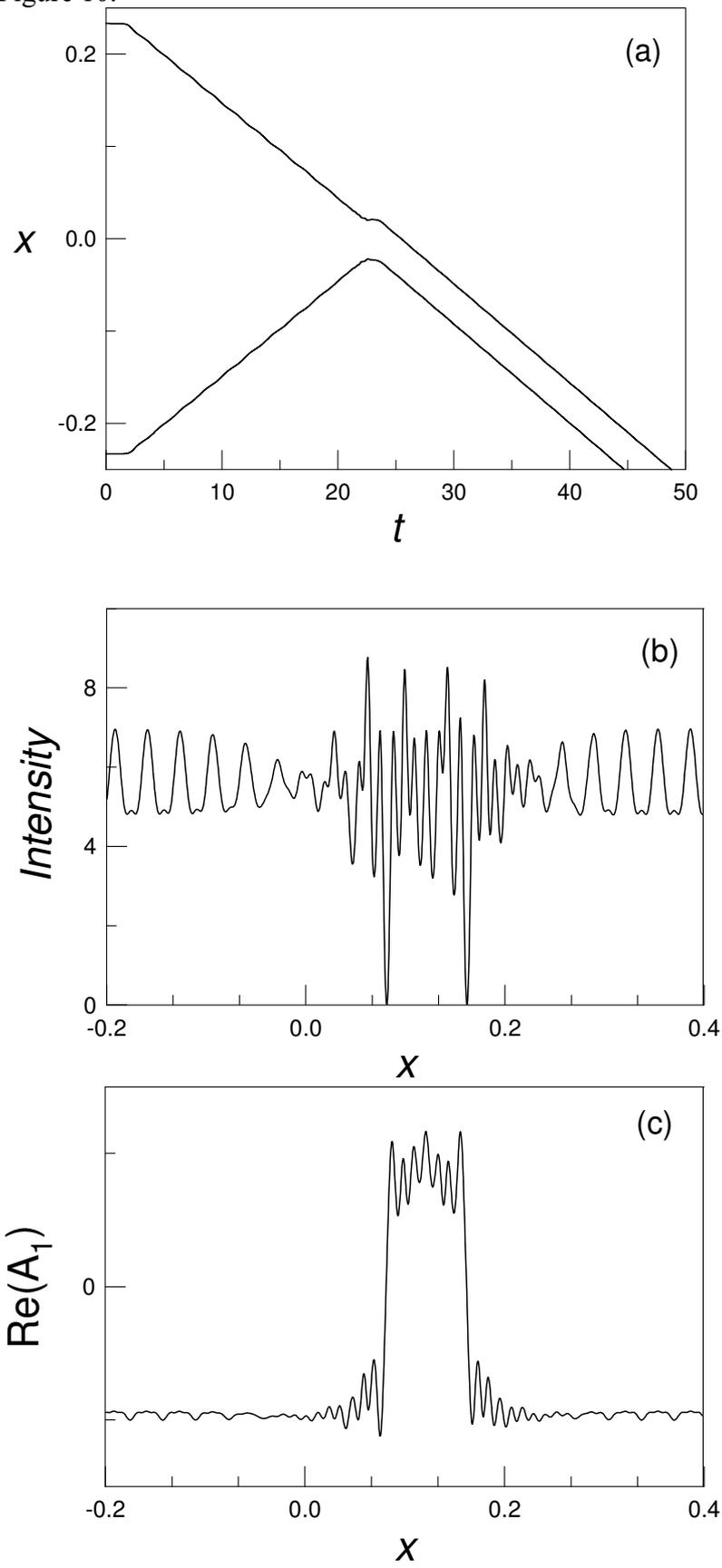
Figure 10. Interaction of two Bloch walls with opposite chirality for
$\Delta=3.25$ and $E=3$ (the rest of parameters as in Fig.1). Intensity (b)
and real part (c) distribution of the bound state.

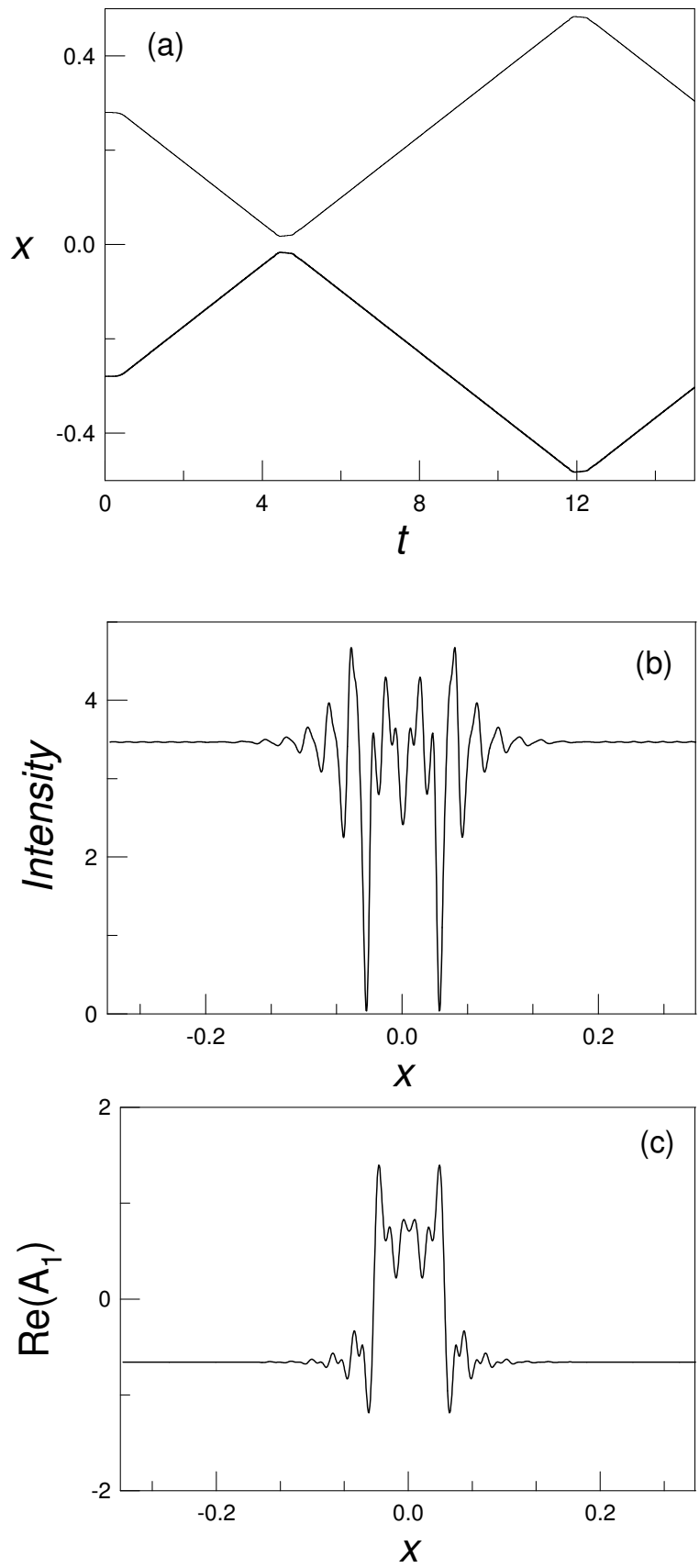
Figure 11. Bouncing of two walls for $\Delta=0.84$ and $E=3.4$ (the rest of
parameters as in Fig.1). Intensity (b) and real part (c) distribution of the
walls during the interaction.


\begin{thebibliography}{99}                                                                                               %


\bibitem {CoulletPRL}P. Coullet, J. Lega, B. Houchmanzadeh and J. Lajzerowicz,
Phys. Rev. Lett. \textbf{65}, 1352 (1990)

\bibitem {Michaelis}D. Michaelis, U. Peschel, F. Lederer, D. V. Skryabin, and
W. J. Firth, Phys. Rev. E \textbf{66}, 066602 (2001)

\bibitem {FrischPRL}T. Frisch, S. Rica, P. Coullet, and J. M. Gilli, Phys.
Rev. Lett. \textbf{72}, 1471 (1994)

\bibitem {HagbergPRE}A. Hagberg and E. Meron, Phys. Rev. E \textbf{48}, 705 (1993)

\bibitem {Perez04}I. P\'{e}rez-Arjona, F. Silva, G. J. de Valc\'{a}rcel, E.
Rold\'{a}n and V. J. S\'{a}nchez-Morcillo, J. Opt. B: Quantum Semiclass. Opt.
\textbf{6}, S361 (2004)

\bibitem {MaxiOL}G. Iz\'{u}s, M. San Miguel and M. Santagiustina, Opt. Lett.
\textbf{25}, 1454 (2000)

\bibitem {Larionova04}Ye. Larionova, U. Peschel, A. Esteban-Mart\'{\i}n, J.
Garc\'{\i}a Monreal and C. O. Weiss, Phys. Rev. A \textbf{69}, 033803 (2004)

\bibitem {Esteban05}A. Esteban-Mart\'{\i}n, V. B. Taranenko, J. Garc\'{\i}a,
G. J. de Valc\'{a}rcel, and E. Rold\'{a}n, arXiv:nlin.PS/0411048

\bibitem {PerezOpEx}I. P\'{e}rez-Arjona, F. Silva, E. Rold\'{a}n, and G. J. de
Valc\'{a}rcel, Opt. Express \textbf{12}, 2130 (2004)

\bibitem {Arecchi99}F. T. Arecchi, S. Boccaletti and P. L. Ramazza, Phys. Rep.
\textbf{318}, 1 (1999)

\bibitem {LL87}L. A. Lugiato and R. Lefever, Phys. Rev. Lett. \textbf{58},
2209 (1987)

\bibitem {Geddes94}J. B. Geddes, J. V. Moloney, E. M. Wright, and W. J. Firth,
Opt. Commun. \textbf{111}, 623(1994)

\bibitem {Hoyuelos98}M. Hoyuelos, P. Colet, M. San Miguel, and D. Walgraef,
Phys. Rev. E \textbf{58}, 2992 (1998)

\bibitem {Gallego00}R. Gallego, M. San Miguel, and R. Toral, Phys. Rev. E
\textbf{61}, 2241 (2000)

\bibitem {Sanchez00a}V. J. S\'{a}nchez-Morcillo, G. J. de Valc\'{a}rcel, and
E. Rold\'{a}n, Opt. Commun. \textbf{173}, 381 (2000)

\bibitem {Sanchez00b}V. J. S\'{a}nchez-Morcillo, I. P\'{e}rez-Arjona, F.
Silva, G. J. de Valc\'{a}rcel, and E. Rold\'{a}n, Opt. Lett. \textbf{25}, 957 (2000)

\bibitem {Perez01}I. P\'{e}rez-Arjona, V. J. S\'{a}nchez-Morcillo, G. J. de
Valc\'{a}rcel, and E. Rold\'{a}n, J. Opt. B: Quantum Semiclass. Opt.
\textbf{3}, S118 (2001)

\bibitem {deValcarcel02}G. J. de Valc\'{a}rcel, I. P\'{e}rez-Arjona, and E.
Rold\'{a}n, Phys. Rev. Lett. \textbf{89}, 164101 (2002)

\bibitem {Longhi}S. Longhi, Physica Scripta \textbf{56}, 611 (1997)

\bibitem {Trillo97}S. Trillo, M. Haelterman, and A. Sheppard, Opt. Lett.
\textbf{22}, 970 (1997)

\bibitem {Kath}W. L. Kath, A. Mecozzi, P. Kumar, and C. G. Goedde, Opt. Lett.
\textbf{19}, 2050 (1994)

\bibitem {Estebanrolls}A. Esteban-Mart\'{\i}n, J. Garc\'{\i}a, E. Rold\'{a}n,
V. B. Taranenko, G. J. de Valc\'{a}rcel, and C. O. Weiss, Phys. Rev. A
\textbf{69}, 033816 (2004)

\bibitem {Boyd}R. W. Boyd, \textit{Nonlinear Optics} (Academic Press, 1992)
\end{thebibliography}
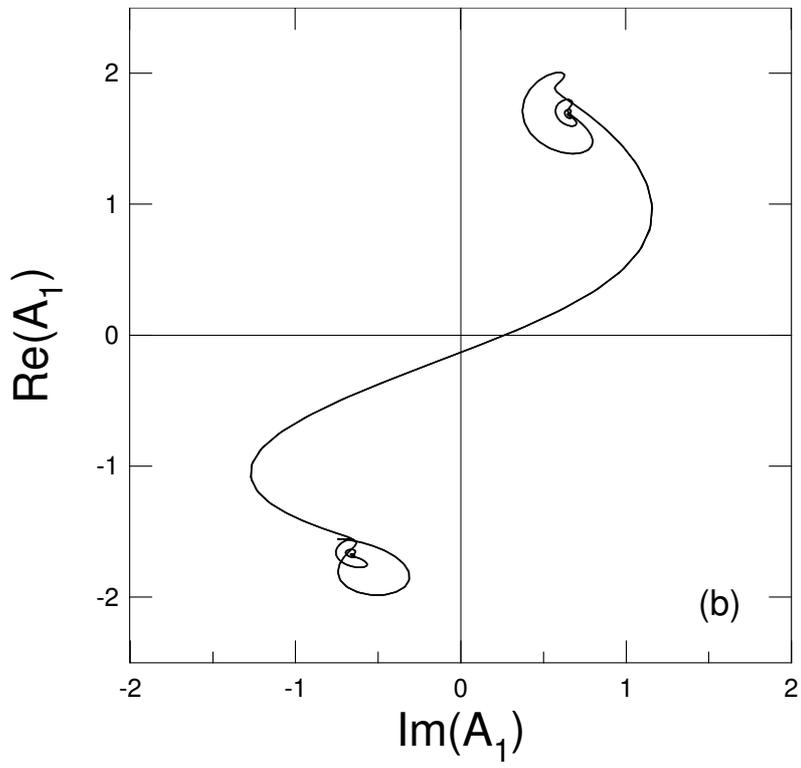
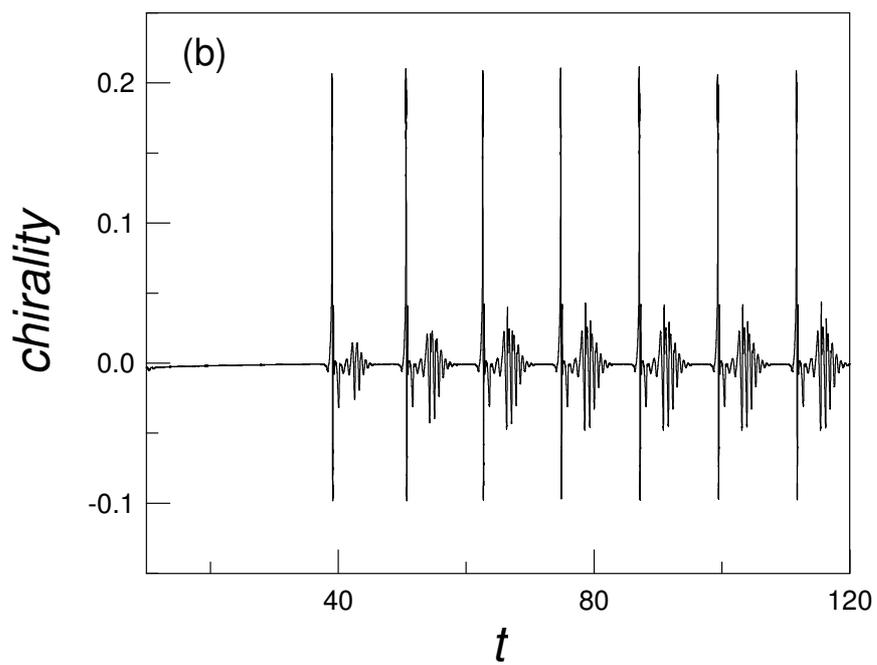
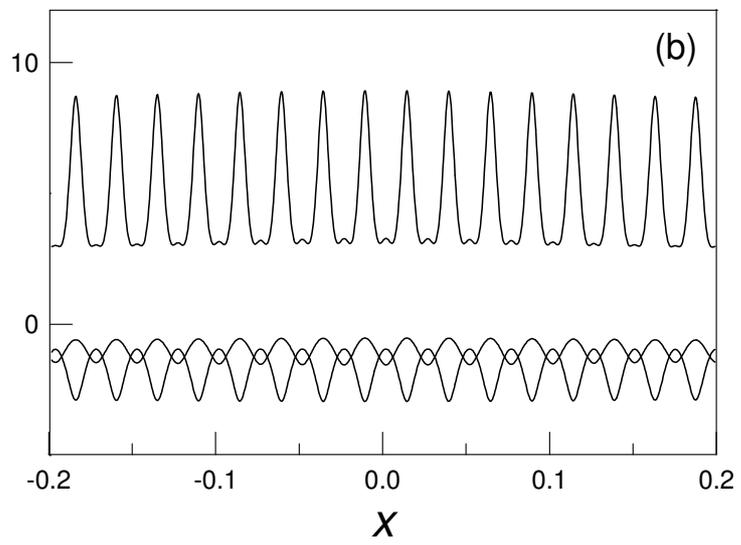
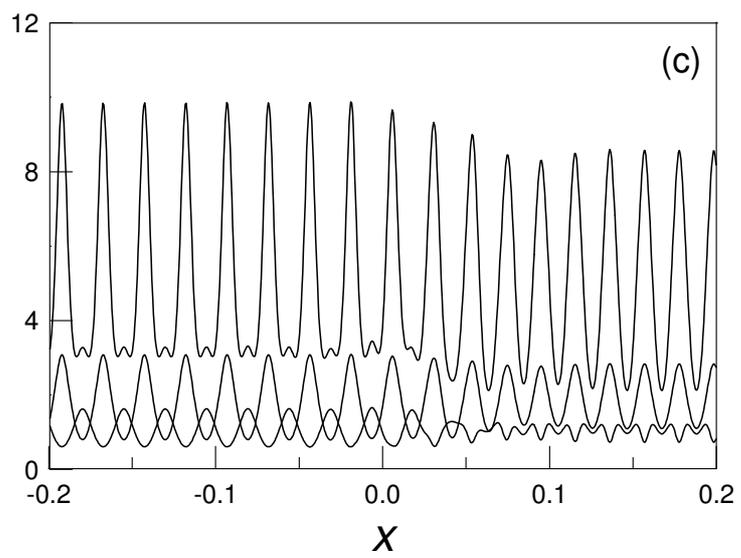
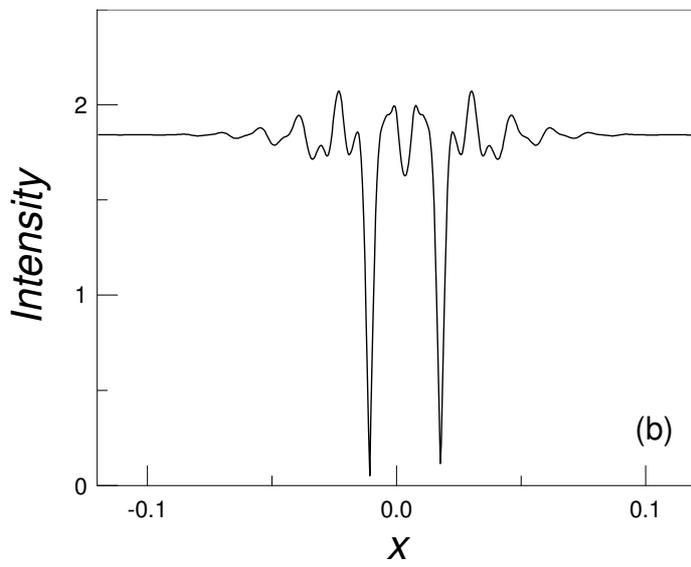
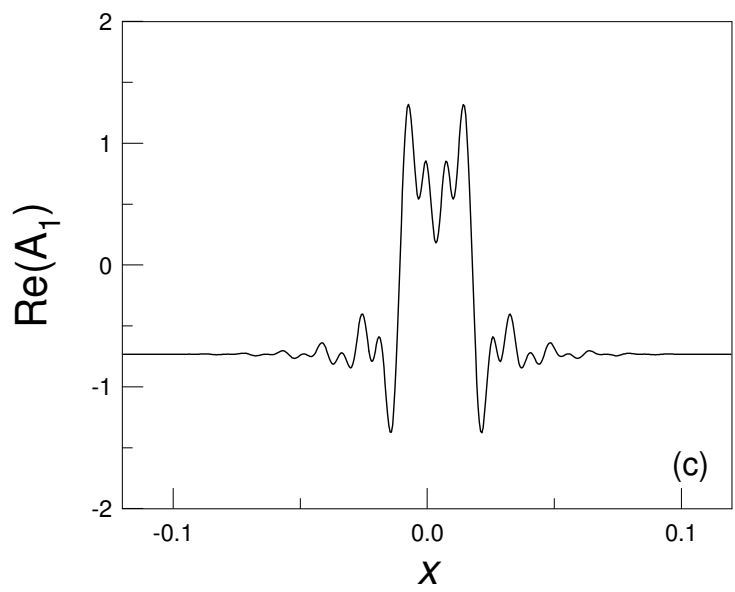
\end{document}